# A PLEA FOR AN UPGRADE TO THE DIGITAL CRAFT OF THE HISTORIAN AND DIGITAL METHODOLOGY FOR DISCOVERING THE PAST

**ABSTRACT**
This essay is about a stricter stance towards the analogue Italian historians' community, to take over spaces in the debate on the historical research field and on scholars' role in turning their craft into digital.
In historical research, we are at a stage of development that is analogous to when the printing press was invented. Amanuensis tried to stop Gutenberg's technology, invoking that more books did not mean that everyone could understand God and Nature. They recognised they were losing control over the culture.
Nowadays, the new digital Renaissance lets computers gain understanding from historical data created and stored in our electronic data stores. However, historians need to upgrade their craft to create that ontology that would drive machines to infer meaning from data and take action based on that meaning.
This essay aims to bid analogue historians assume that digitisation is the first step to creating historical heritage based on the new language of Science: Computer Science.
As we know, Humanities disciplines cannot easily be encapsulated in a few understandable numbers and names. However, historians must boost Artificial Intelligence (such as Transkribus) and Neural Networks to let the Machine infer meaning from the digitised historical primary source and become the most powerful tool to help historians understand what happened in the Past. Historians – collaborating with data scientists, expert annotators, librarians, archivists, and others, who are crucial to the successful management of digital data collection – have to create the primary ontology, starting from coding manuscripts into digital text, as the Biscari Archive (Italy) study case.

**KEYWORD**
Biscari Archive, Transkribus, historical digital heritage, HomoLogatus, Biotecnocenosi.

## 1. SOME CONSIDERATION THROUGH TIME.

In our World, where every single day remembers that digital innovation is a way of living, historians need to upgrade their skills to release the digital version of "their craft".
Nevertheless, analogue historians appear to be stifling an opportunity for a dialogue that seeks to build a link between History and ITC, if it was true 60 years ago and even more so nowadays.
«No meet this bar», writes Milligan in 2019 (Milligan 2019). Historians concern with their theological and scientific ideas because of a lack of innovation in the Statute of History.
From the historiographic theories of the nineteenth century, which comprise the basis of Historical Science, there was no innovation of its epistemic approach, which is still based on the concept of "research without a laboratory" based on the scholar's intuition (Spina 2022).
Historians' ideological approach is aimed at preserving their view in the science community because they do not have a suitable set of tools to dissect and understand digital dimensions and how ITC could improve their craft.
In 1821, Wilhelm von Humboldt published *Ueber die Aufgabe des Geschichtsschreibers* (Humboldt 1990), exposing historians' roles and research aims. Nevertheless, all definitions show how accomplished constructing an epistemology of History is. The masterwork by Marc Bloch (Bloch 1963) (*The Historian's Craft*) is evidence of a constant determination to build a systematic structure for the History Statute, which allowed historians to place themselves in the field of methodological sciences.
Taking a cue from Leopold von Ranke, the most critical historian of the 19th century (the one who gives shape to the scientific figure of the "professional historian", stating that the historians' task is to "show what really happened" or "how things are"), historians have to write about the Truth, the whole Truth, analogue and digital (Ranke e Ramonat 2010). So, he cannot close dialogue with computer technology insofar as the latter has the tools to bring historical Knowledge to a state of objectivity, which was impossible to achieve. On the other hand, it represents Science's linguistic system. ITC is the communication code that guarantees the multidisciplinary approach Historical Science requires.
As Mathematics, during the 18th, 19th, and 20th centuries was the language of Natural Science, Computer Science created the linguistic bridge between all fields of Knowledge, opening Humanities with the possibility of an interdisciplinary approach and creating an "epistemological laboratory" that lets humanists test their theory (Spina 2022).
Word and expression, in fact, thanks to Data Mining, Machine Learning, and other tools, can be analysed as numbers and mathematical functions. For a digital historian, Computer Science is the laboratory to experience and discover what happens in the Past.
For instance, thanks to Computational Linguistics and its close link between Language studies and Historical Science, Cesare Vetter demonstrates (while considering the purely evidentiary character of the analyses) that applying ITC tools to primary historical sources is possible to increase Knowledge of many events of the Past. Considering the History of Ideas, Vetter describes the concept of "Freedom" that guided revolutionaries, marking all the events of the French Revolution. Also, Vetter tested how "Freedom" suffered an evolution that brought Robespierre to those ideological positions that arose in the Terror period (Vetter 2005; 2007; 2018).
With computer tools, it was possible for Vetter to quickly specify the juncture for Robespierre, in which "*adversaires*" become "*ennemis*". Also, thanks to Computational Linguistics, it was possible to recognise when the expression "*Ennemis du*

*Peuple*" appears for the first time and when Robespierre qualifies his opponents with slanderous words such as "*insectes*" and "*monstras*". Moreover, text analysis shows how hurtful language aims to delegitimise dissidents, creating suitable conditions for dictatorships.

Digitising a corpus of historical documents allows historians to verify their intuitions and a more excellent intellectual honesty of the study results. Because the relation between interpretative paradigm and empirical material becomes transparent and easily verifiable, the researcher can move with "freedom and curiosity, and avoid forcing the texts".

## 2. A BALANCE OF IDEOLOGICAL POSITION?

Historians build their craft thanks to Memory, Archives, and all those human expressions that testify to the Past. Computers, software, and algorithms are, as documents and monuments, "jotted-down facts" preserved on different media than traditional paper. Nevertheless, ITCs are testimony to Man who makes them.

Towards the computer, analogue historians express an intense pessimism (Judt 1979) in an attempt not to lose political control of the historical research field. So, the historians' community limits the path of this research field in the digital environment. They force it to reject considering computer technology, on the one hand, as "events" to debate and, on the other hand, the main tools that can analyse and draw much more information from archival sources than the analogue close reading approach (Rowland 1991).

The reason for close-mindedness towards applying IT tools, such as Data Mining or Machine Learning, is determined by the evident unpreparedness of scholars, such as their inability to create databases, analyse data, and apply algorithms and Artificial Intelligence.

However, in a World where the man-computer bond is increasingly vital, historians try to escape this "biotecnocenosi" (Spina 2022) that distinguishes this Era. For the analogue historian, a computer is:

1. An information collector (mainly in the Excel file).
2. A replacement of the typewriter.
3. The way to consult archival document registers.
4. In an exceptional line, to display digitised sources.

For instance, there is much talk about digitisation and digitalisation in the analogue historical community. Still, the two terms are attributed a wrong meaning by scholars: "digitisation" is like "taking a photo". So, to an analogue historian, "taking a photo" is one of the assumptions to build the Statute of Digital History. Most scholars believe the digital way of the historical methodology is to scan or take photos of some archival document and upload them to a database (built, the latter, by someone else).

For Digital Historians, however, the computer is the best tool to study man's actions, complex societies, and all those dynamics that have marked significant and minor historical events.

Algorithms and artificial intelligence can mine information, infer meaning, and trace new research paths, showing a scholar new resources and data.

The rapid development of Computational Analysis, methods, visualisation, and interpretation of "linguistic meanings" (the binding force of all large and small communities) can boost historical research, which can look at men as elements that create correlations between them. In this case, the latter can be studied and analysed by applying the same Physical and Natural Sciences methodologies.

Information Technology is based on Artificial Intelligence, data, algorithms, and analysis. So, to apply to ITC on their ground of scholarship, other Sciences need to start with "coding" and "formalisation". Any Science that seeks multidisciplinary cooperation must "codify" its Statutes, Paradigms, and laws in complex categories and relations (ontology), which the computer will use to process information.

It becomes necessary for a digital historian to know the ITC system to build an ontology that lets Artificial Intelligence create new Knowledge to satisfy scholars' queries and questions.

In a potential future where the digital storing step of the documentary and cultural heritage of all Mankind comes to fulfilment, historians must provide their skills to build a network – taking as a model Wu-Dao, the most extensive Chinese neural network –, that could connect and deliver new Knowledge to scholars. Historians must build a new "*heimat*" where the link between humans and computers is systemic (Jouman Hajjar 2021; Pati 2021; Zhavoronkov 2021).

So far, historians are connected people to the Internet. We need to create an "Information Exchange System" (IES) where historians become *HomoLogatus*, the central node of the digital version of historical methodology (Spina 2022).

In 1968, Le Roy Ladurie, in his essay *La fin des érudits* (Le Roy Ladurie 1968), examining the progress of the historical quantitative approach, which characterised the Historiography of most of the 20th century, expressed his idea on the necessary computer training of historians and historiographers. Only understanding Computer Science and Information Technology would have guaranteed historians to comprehend the Hyperlearning Revolution, which Perelman (*School's Out*) spoke about in 1992 (Perelman 1992).

Thirty years later, the tone could not be more different. Most historians do not consider computers as the cause of a historiographical revolution, unlike Le Roy Ladurie. For the latter, instead, ITC has changed the course of History since the 1930s (Le Roy Ladurie 1973). Just think of the research conducted, for instance, by Aydelotte, who, thanks to the use of computer systems, succeeded in reconstructing, with a scalogram analysis, the voting models of the English parliament in 1840 (Aydelotte 1963).

Historians, however, do not perceive this real opportunity provided by Artificial Intelligence. Nevertheless, their assumptions and ideas fail to achieve a significant settlement.

Most Historians, like Rosenberg, believe that humanists need to train a new historian. In comparison, analogue historians strive to consider computers a knowledge tool (Rosenberg 2003).

To this, add Dougherty and Nawrotzki's idea that thanks to technology will, historians could move between events, add or modify facts that have been previously described, and look at new interpretative variables (Dougherty e Nawrotzki 2016). Moreover, the digital approach to historical scholarship could bridge the gap caused by the generational divide. The young researcher will not visit Archives but will demand access from the Web and, above all, ask for access to digital editions and transcripts of primary sources.

Therefore, an upgrade of the craft and skill of scholars will be necessary for future research in the field of History. As Brügger explains in his book "General", digital historians will have to look at the Web, on the one hand, as a gate of access to the documentation that archives will digitise, and on the other hand, as a historical source (Brügger 2012). For instance, the History of online discussion helps to understand the History of the Internet itself.

Historians live in their contemporary age, so, as Chabod states, like the mass media (Chabod 1969), ITC is the most important historical source to understand mass psychology and how a man lives in the 21st century, the age of digital turn. Scholars have the duty to grasp and explain the meanings of what surrounds them. As stated by Benedetto Croce, historical scholarship aims to explain the historians' contemporary scenario (Croce 2002). History fills the Present, and analogue historians must leave their obsolete points of view behind and spread their minds to digital methodology and a new concept and kind of primary sources, such as email or tweets.

We have to assume that, just like the diary of some Great War soldier had no historical relevance at that time, today, the same consideration is reserved for a social-network post, a comment, or a tweet. Only the distance in time and the concrete desire to understand what happened somewhere sometimes make a personal diary a primary historical resource. The soldier does not consider what he is writing from a historical point of view. He wants to catalyse his feelings, certainly not for his desire to be involved in history writing.

Nowadays, the billions of posts and comments concerning Covid19 and the war in Ukraine do not currently have historical importance. However, in time, historians will look back on this data as the essential source for understanding what we have experienced today.

## 3. ITC AND HISTORIANS

"Number" is the assumption on which digitisation is based: everything that can be quantified can then be tried, and if it can be processed, it can be digitised.

A manuscript photoproduction has much information that the most attentive reader can find as he can read its contents. In a computer-conducted analysis, the same image is a set of data that excludes those that can only be obtained through human interpretation and intuition. So, to narrow this gap, scholars dealing with digitisation processes have been trying to understand how to create a computable text. Computational Linguistics is based on this aim. A text written in the digital environment lets us conduct simple textual analysis, but a native-analogue text photographed is invisible to the Machine. Those signs that we know to be words, for the computer, are a set of pixels that have, for example, a colour code (such as the "hexadecimal") different from the others present in the same image file (the "black" of the Ink is #000000, brown is #654321, the red is #FF0000, and so on).

What it is necessary to do, is to "inform" the Machine that that pixel set is, indeed, a word and, after a process of comparison with models of words transcribed by man (with which the computer is formed to recognise characters), ask the computer to transcribe those signs.

We have to consider that, beyond the challenge of "formalisation of meanings" and the pairing between the way of human processing and that of the Machine, it is true that, to boost the possibility of processing the Big Data of History, we need to convert (transcribe) the whole archival manuscripts heritage into digital texts.

Therefore, an instrument is needed to return the digital format (.txt) of writing preserved on paper support, as Dimond was already attended in 1957 (Dimond 1957).

From that date, over these decades, several projects aimed at the automation of transcription (Handwritten Text Recognition – HTR) arose (Spina 2022).

The most ambitious project, however, is Transkribus. Proposed between 2015 and 2019, nowadays, this AI is developed by the READ Cooperative (Recognition and Enrichment of Archival Documents) and funded by Horizon 2020 to develop and promote a «functioning online research infrastructure where new technologies can feed innovation in archival research» (Erwin 2020; Kahle et al. 2017; Milioni 2020; Muehlberger et al. 2019).

The process of mass digitisation of historical archival resources is a work that will not bring an increase in the historical research field without the transcription of the whole text. No algorithm or computer-assisted analysis could provide innovative achievements if the only information concerns limited metadata, such as author, date, and place. These details are not the Big Data of History. If we want much more from Past's documents, we must digitise every single word, let every AI read them, and provide detailed information to the historian's query.

The role of the Transkribus teams is to open the possibility of a mass automated transcription and boost access to collections, allowing users to quickly and efficiently pinpoint topics, words, people, places, and events in documents. An entire corpus of transcribed documents lets historians change their perspective, understand a new context, and find new ways to his historical matter.

Of course, scholars have to deal with their skills, while, on the other hand, the CER (Character Error Rate) of Transkribus models must be lowered to a level that allows historians to have correct transcriptions. Although, the neglect of a thousand documents, with an inevitable error in recognised text, is preferable to an analogue project in which the manual transcription is limited to a few pages.

Moreover, we must consider that a thousand and more transcribed documents could be inserted into a database, which facilitates the work of thematic selection by the scholar.

However, what is the role of historians? As stated in our title, they must release the upgrade to the digital version of their craft and skills. No matter what "release" we should start with and what we should come up with (2.0, 3.0, 4.0), state of the art of upgrading process, historians must focus on the goal of being the only ones to trace the path of re-ontologising scholarship in all fields.

It is the task of historians to «meet the bar» and overcome what scholars are experiencing as the "digital storing step" of the whole of Mankind's cultural and historical heritage. The "coding and process step" is still far away for now, but not, for this reason, all digitisation projects implemented by many archives worldwide are limited to storing historical heritage in some database. The realisation of an overarching ontology to process data remains the primary goal. The Turing Machine requires, in fact, something very different from a photographed document. Artificial Intelligence needs correctly coded data (straightforward and formalised) «apt to formulate an algorithm», as Tito Orlandi stated (Orlandi 2021).

Digital Historians have, therefore, the need to ask themselves which ontology to create, as a complex of concepts, categories and relationships must contribute to making that the computer can process and "explain" historical questions to get meaningful information to support Our research hypotheses on an event of the Past.

Digital historians, as well as all digital humanists, must focus on the computation step and the possibility of filling the gap between rationality and intuition through tools that will have to improve more and more in the ability to understand the arbitrariness of the characters and symbols.

"Deciding what they mean" is not the prerogative nowadays of a computer. Intuition arises from the possibility but is based on human arbitrariness that prevents the Machine from deciding, as it is not computable. This function belongs to historians, of course.

Text encoding, XML, HTR and all ITC tools on which digital humanities are based still fail to provide computer systems with the possibility of understanding and intuition. For this reason, it is necessary to train a historian who, by digitalising his craft, manages to be a medium of the need for research to codify the historical heritage in a digital computable estate.

As Niel Bohr brings classic Physics to the quantum (explaining that without a conceptual transformation, there will be impossible limits between Science and scholars), historians, in a *poietic* path, need to translate their know-how and skills to implement Gordon Bell's fourth paradigm.

So, this essay is an enduring appeal to reflect on the near future historical research, which will be digital *tout court*. Digitisation is re-ontologising the World, but humans remain at the centre of language; it does so through tools such as Transkribus, dhSegment, Vojant, Recogito, Palladio, Vistorian, and all those that will be developed in the future. The technologies will have to contribute to building a digital heritage that is not simply filing, recurrence, layout analysis, and transcription. Historians, data scientists, librarians and archivists must ensure that the whole historical heritage is codified and formalised to allow the Turing Machine to elaborate and deduce valid meanings for itself (Machine Learning) and the historians' scientific community. For this, the latter must become the only matrix of the foundation of that ontology that will be the required mechanism of the semantic Web and of an AI that progressively enhances its algorithms.

Historians will need to understand the "aim" of the Turing Machine, its "inner" mechanism, and the need to assemble a knowledge that can be remodelled in its linguistics, its meanings, in its representations, to make the analogue and digital World intelligible for each other.

Historical Science and historians must, therefore, play the role of the arch of that ontology/dimension/system of coexistence of meanings (analogical and digital). Historians are *HomoLogatus*, but with the straightforward task of realising the "biotecnocenosi", which is a state of balance of these systems through a linguistic system that allows the Machine to discern human Knowledge. ITC is part of the "cenosi[1]" and, therefore, an involved party that brings its possibilities in scholarship for that Hegelian historiographical fullness, «which must accommodate all the speeches ... which are the principles of time» (Hegel 2022).

This assumption, although traumatic, is only "the act of starting" (Vico 2019).

We live in a time when we have a very sophisticated means of communication, but the latter has become very difficult between individuals and groups. In all its issues, the public debate is fragmented and polarised by post-truth and fake news. Meanwhile, History has become an open field, and network communication is a whole industry based on Silicon Valley's ideology.

From all this, by making the interpretation and conceptual system, historians pull out. They should, having the tools, protect the speech on the Truth.

There is a severe danger of the deliberate decision of analogue historians to fix and not wind up the debates, cutting off their role in the scientific community. This willingness needlessly creates a self-referential historical community; on the other hand, it allows dissimilar specialists to use different approaches to speak about History.

---

[1] *Cenòsi* s. f. [from Gr. κοίνωσις «union»; cf. ceno-]. - In Biology, is the set of plant (phytocoenosis) and animal (zoocenosis) species living in the same environment. Synon. of *biocenosis*.

If historians renounce to speak publicly about History due to their poor digital technical skills, Communication Companies will also have the opportunity to monopolise Historical Knowledge. ITC Companies, as owners of the data, could discuss everything with a high competency rate.

We know what a significant function amanuensis held in preserving and communicating Knowledge. Nevertheless, Gutenberg stopped their thousand-year activity. After the typographical turn of the 16th century, nobody (especially nowadays) asks a publisher to write copies of his book manually. In the same way, the analogue historian, who opposes all the criticisms and theoretical defences, will be overcome by the digital version of his craft.

We do not fund in the analogue. It is a fact. Furthermore, we do not support analogue research that wants to maintain a unique idea and performance, even if the use of ITC. A historian who accesses the Internet to look for primary historical sources in the online registers of the archives is not a digital scholar but simply a connected researcher. His set of minds is, however, analogue. So, no one will fund a researcher who does not have computer skills. An analogue historian has to avoid the carriage's fate. Investment in rail and train development could not but be convenient. We can power a locomotive, not a horse. So, the Industrial Revolution shaped the "analogue means of transport" destiny.

So, nowadays, in the hyper-technological World, the terms "carriages" and "horses" recall a millennia-old era, but they denote something different.

In the same way, the digital historian's ITC skills will allow him to stem the authoritarian drift of Communication Companies and regain the function of the only interpreter of past events.

## 4. THE BISCARI ARCHIVE. AN INSTANCE OF A DIGITAL HISTORICAL WORKFLOW.

While tanks to ITC we can organise and displays data, historians must infer meanings from that organised information, an assumption that shows scholars' short-sighted attitude and the immovable division between mining and analysis established by Johannes Kepler when he discovered the laws of planetary motion.

However, can the Machine give a speech on History in an Era where we can overcome all divisions between technology and human interpretation?

That crucial question leads us to conduct a study on some primary historical sources to demonstrate if Artificial Intelligence and algorithms could infer or show new meanings after analysing some archival documents.

The archive of the Paternò Castello family, Princeps of Biscari (today, Acate – RG, Italy), preserved in the State Archive of Catania (Italy), is a unique complex of computable historical data.

Among the 2,000 thousand stored folders, which document political and economic reforms, controversies and disputes, and government decisions, the "Correspondence" section consists of about 42,493 sheets. Inside folder number 1642, we find 591 sheets that constitute 366 letters and one manuscript of Emile Rousseau.

Although it would have been appropriate to choose computable sources in a computational environment, the heterogeneity of data stored in folder 1642 was crucial for demonstrating the feasibility of the application of the ITC tool to describe how the Paternò Castello family built their mighty.

All data about people and places has been stored in a database created with Claris Filemaker Pro 19 software.

Every piece of information is indexed, and the 367 records have been compiled thanks to two relational substructures where data on people and places have been stored.

Each record collects the images of the letter and the related metadata. Through an import process from the two substructures, the form is compiled, which thus becomes a "container" of related information: sender, consignee, date and place of issue, city of destination, subjects and other sites named in the documents, register and keywords.

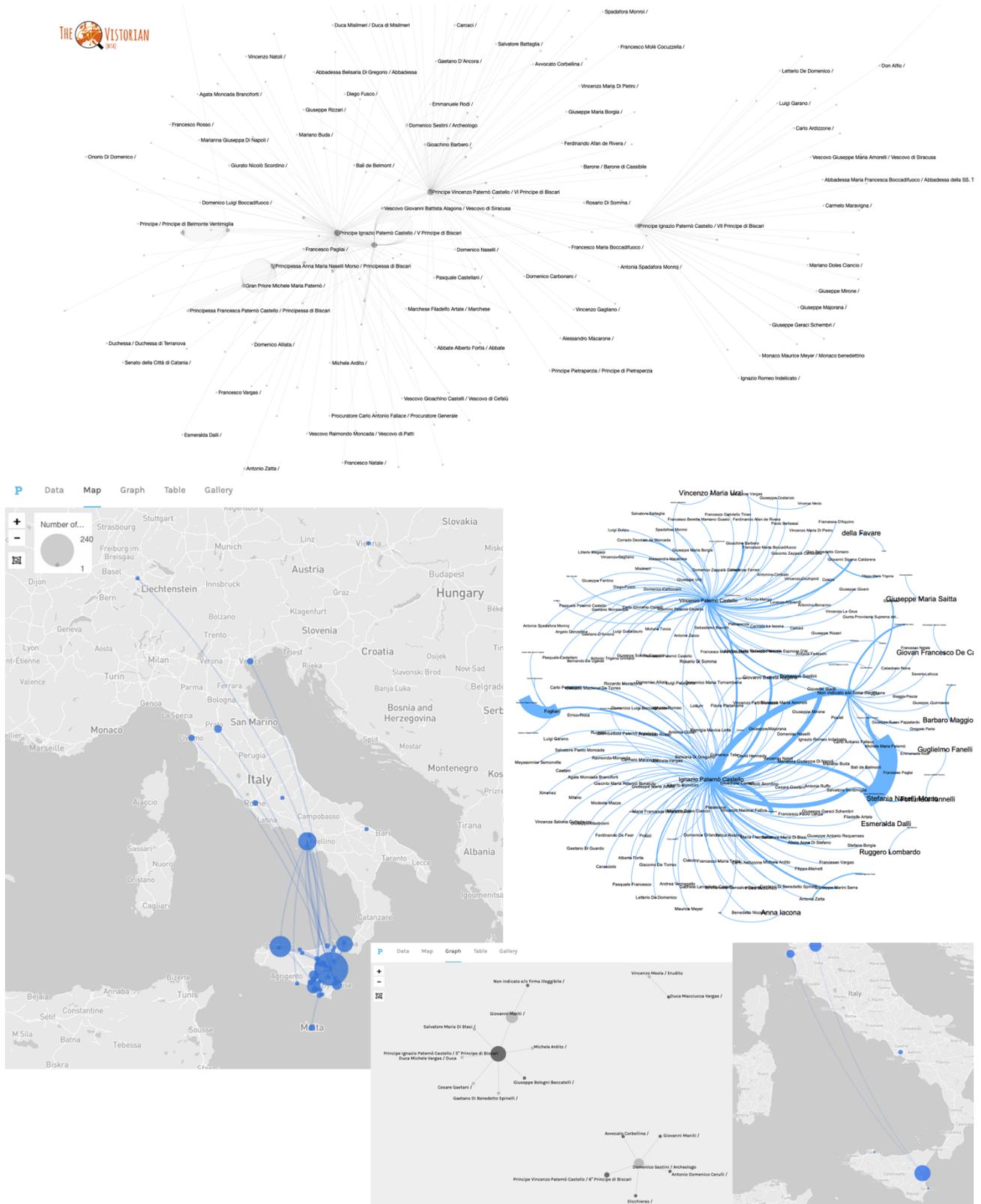

*Figure 1. Sender-Consignee-Place Network.*

Thanks to this function, it is possible to export data in different file formats (for instance, CSV), allowing scholars to conduct further analyses. For example, a "sender-consignee" CSV file was exported to applicate "The Vistorian" tool. In comparison, the "Palladio" tool was applied to analyse a different CSV file, where the geo-coordinates let visualise how Biscari's network in space extended.

Thanks to Palladio, it was possible to recognise that the important centres of the power management of the prince of Biscari were Catania, Palermo, and Naples, but their influence reached Zurich, Vienna, and Malta; demonstrating how essential the Princes and Princesses of Biscari's policy idea of was in Europe.

The epic of the Paternò Castello began in 1623 when Kings Filippo 4th of Spain granted Agatino Paternò Castello the title of "Prince". From this date, Biscari's leverage reached its maximum expression with the fifth, sixth and seventh titles.

What can we infer from the 367 letters? An (analogue) close reading approach leads a historian to spend several months reading and transcribing the preliminary information from the documents. While in the digital environment, historians can use a specific tool to HTR historical documents. So, it was decided to upload all the letters' images to the Transkribus server to transcribe and quickly create the digital corpus of the correspondences. The Keyphrase Digger tool analysed the latter to extract more complex information.

The outcome was encouraging. The traditional historiography describes Ignazio as a patron of the arts, but scholars neglect many data. The correspondence content analysis shows more information than the analogue approach.

Ignazio loved surrounding himself with scientists and scholars who could help him reconstruct Catania after the Etna earthquake of 1693.

However, the network analysis demonstrates that the management of political affairs was entrusted to the wife of Ignazio, Princess Anna Morso.

Thanks to the Gephi tool, it was possible to view a peculiarity: inside folder 1642, 28 letters were written by the Prior of Messina, Michele Maria Paternò.

Thanks to the Keyphrase Digger tool, it emerged that Michele is linked to the family of Biscari (M. C. Calabrese 2016), especially with Princess Anna Morso, a constant consignee of the letters, which confirms to be a key figure in the political and administrative relations that the family builds on the territory of the Kingdom.

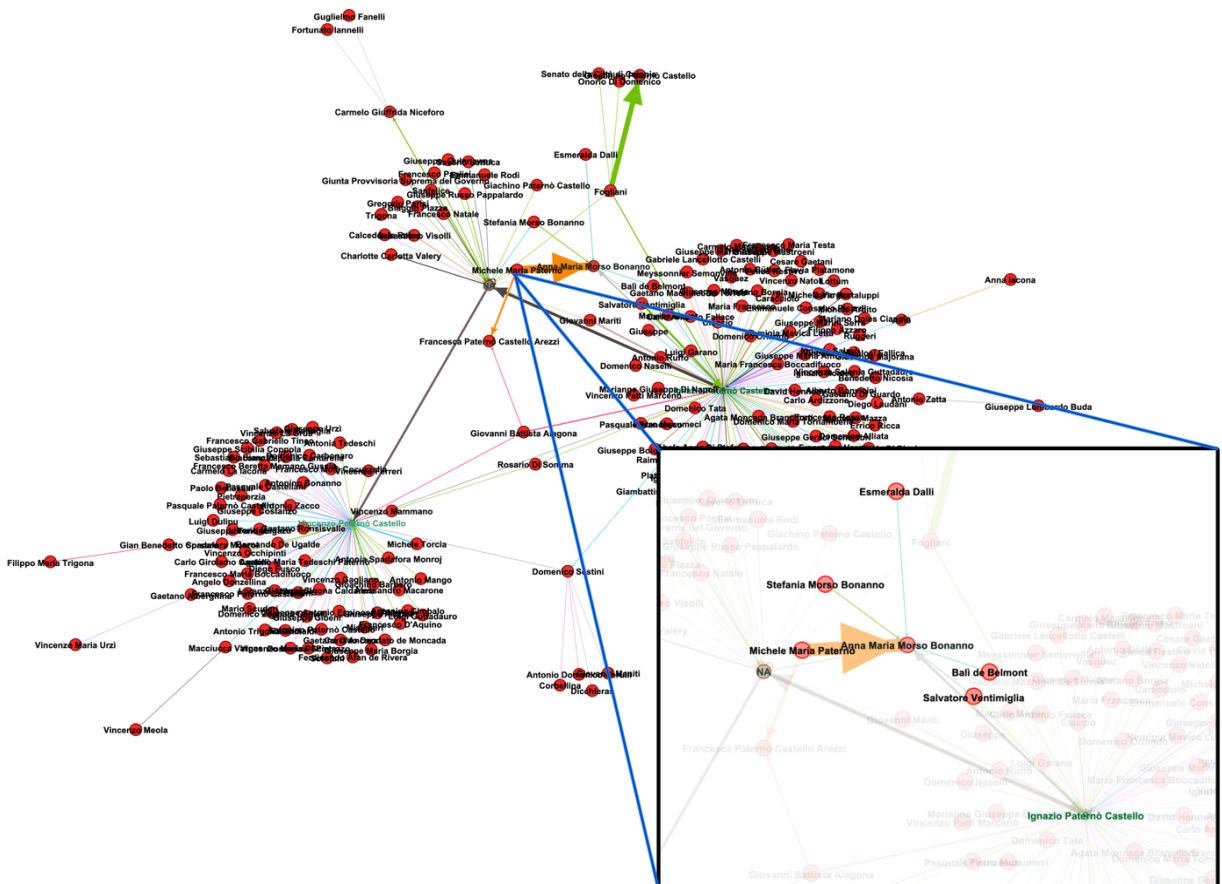

*Figure 2. Correspondence between Michele Maria Paternò and Anna Maria Morso e Bonanno.*

While Ignazio deals with Arts, Anna has political power, contrary to what historiography claims (G. Calabrese 2003; Emanuele 1985; Muscolino 2015; Sestini e Giorgi 1787). As we can infer from the content analysis, Anna wants to be a central node of the political system of the Naples Kindom. In order to rule, she needs to know what happens at the courts of Palermo and Naples; in this case, Michele, and his network, are the right man in a central place, in a historical moment in which France is the tiebreaker between Austria and Spain, in an attempt to bring Sicily under Habsburg influence. In the second part of the eighteenth century, in fact, international relations began to shift, leading to the policy of the Kingdom of Naples, where Bernardo Tanucci tried to stem the not-very-careful policy of King Ferdinand IV.

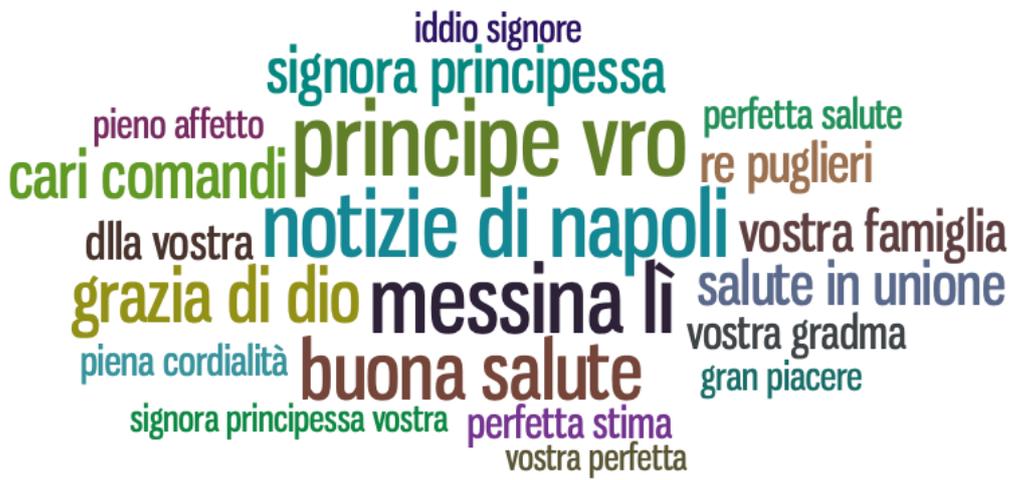

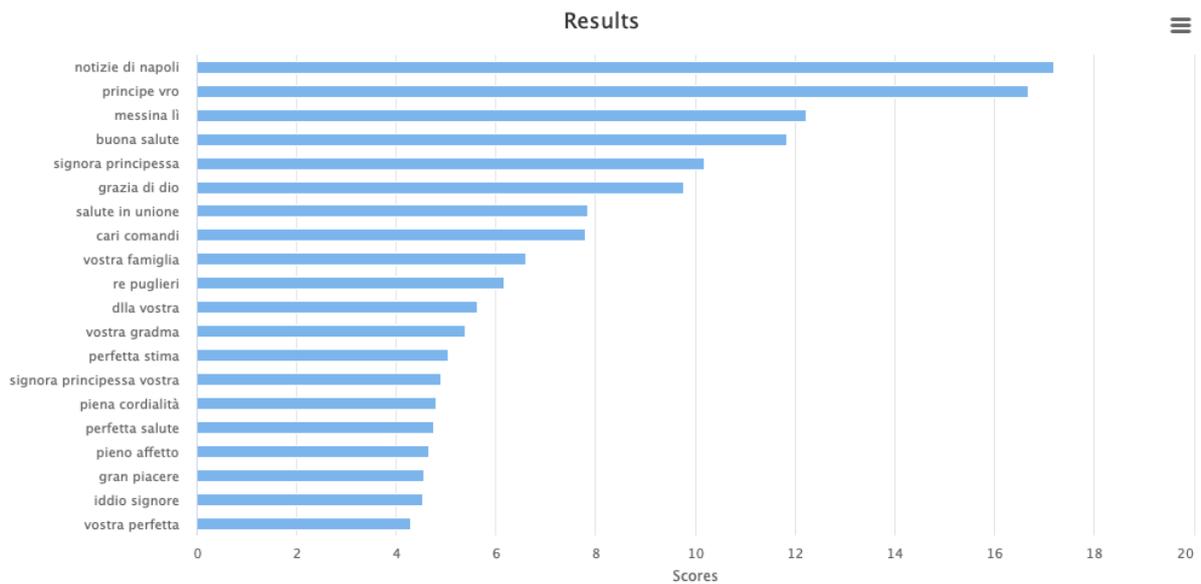

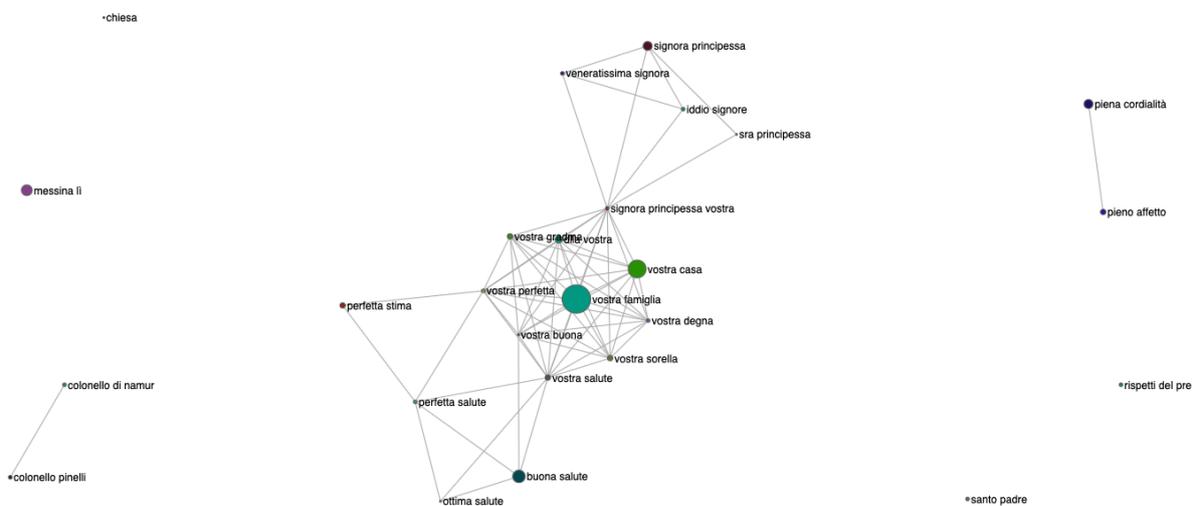

*Figure 3. Keyphrase Digger outcome.*

Let us continue with the matter.

The network of scholars mainly extended to the territory of the Italian peninsula. Every scientist and philosopher was reached to help Prince Ignazio identify historical and artistic wealth to enrich the city of Catania and, in general, to deepen the History of Sicily. In the text of the letters, we find several themes. They range from analysing engravings on ancient marble to the exact site of statues and columns in cities, from medications to wild wheat, to disputes on the foundations and destruction of Sicilian towns.

> Lettera
> Scritta a Gio: Mariti mio Cugino in Firenze Sopra quei quattro Donarj fittili, rappresentanti alcune Sacerdotesse con una pianchetta in mano, trasmessili in dono da S. E. il Sig.r Ppe di Biscari

> e ne fecero intorno all'antica Città di Camerina, e specialm.e Pindaro Tucidide, Polybio, Diodoro, e tanti altri [...], che

> Per mezzo del deg.mo vro Sig.r Cugino D.n Gio: Mariti, mi perviene da Firenze un' eruditissima vra lettera a me diretta, continente la spiegaz.ne di certi flagelli usati dagli Antichi, Plumbate appellati, de' quali me ne accludete il disegno, appunto come rattrovansi in Cotesto scelt.mo [...] Ill.mo Prle Principe di Biscari, del qual Museo siete Voi degno

> Soggiungendo di più, che non solo si vedea nascere il frumento salvatico da se stesso nell' agro Leontino

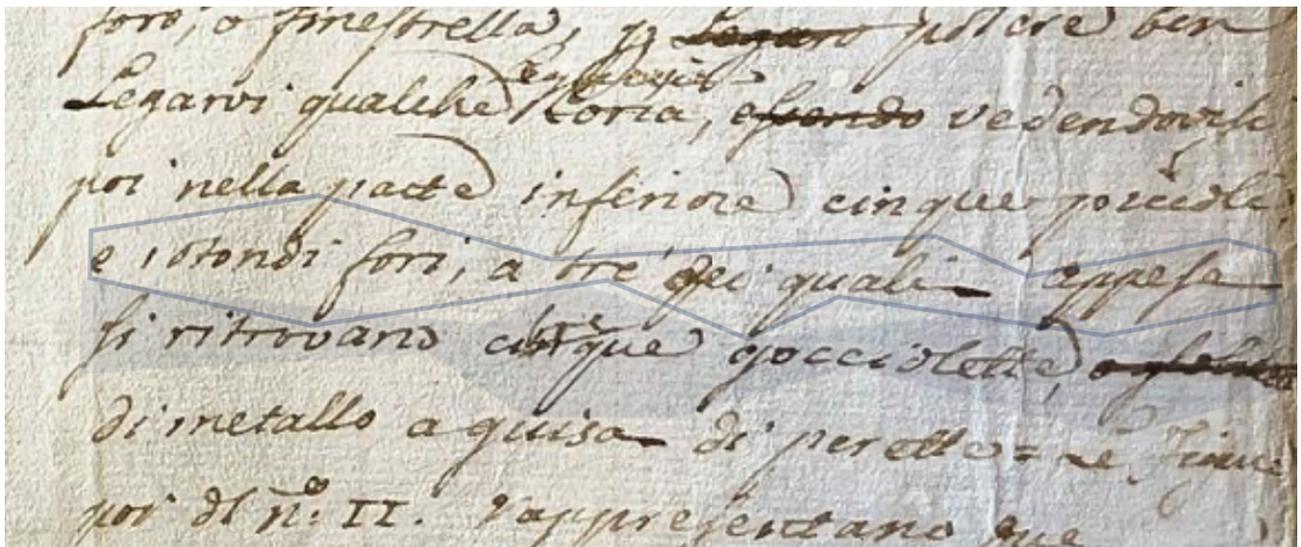
*Figure 4. Text segmentation.*

The letters are most of the copies the consignee sent to Prince Ignazio, who was interested in correspondence among scholars regarding specific issues. Alternatively, they were epistolary exchanges concerning other people to whom the prince wrote by an interposed person. For instance, the relationship between the Prince and Domenico Sestini, the archaeologist Ignazio wanted as curator of his museum. The connection between the two people is driven, in fact, by Giovanni Mariti, cousin and teacher of Domenico, to whom he will have the task of managing this correspondence, informing the prince about the scholar's activities, about his movements and, above all, of his relations with the Middle East.
Mariti is, therefore, the pivot of friendship between Sestini and Ignazio. Furthermore, it is always Mariti who will take care of the associations between Ignazio and the Academy of Georgofili.
The network is also made up of other fellows and scholars, whose presence is proof of the various cultural interests of the Prince of Biscari: Cesare Gaetani, Salvatore Maria di Blasi, Gaetano by Benedetto Spinelli, Michele Ardito, Francesco Vargas, Duke Michele Vargas, Giuseppe Bologni Beccatelli.
The digital historical approach to archival heritage lets historians shows their results without writing. The digital historian is no longer a scholar who must necessarily write.
Indeed, among the most helpful and frequently used approaches, the Historical Network Analysis (HNA) is the best to give meaning to information, create new Knowledge, and refute theories on past events.
The HNA is based on the assumption that man builds relationships with others similar and with the surrounding space. Thanks to Network Analysis, it is possible to understand how a man of the Past lived his life, how he related with other men and where he moved.
Nevertheless, now, we need to answer the question: Machine showed new data that led historians to change their Knowledge about one of the most influential families of Sicily; is this the case of a computer-assisted historical analysis? Or is it the case of a human-assisted study that lets the Machine show the authentic network on which the Prince of Biscari builds his mighty? Can a machine start the historical workflow (digitisation, transcription, analysis, meaning infer, visualisation) to unassisted conduct a study?
Is this the way to Gordon Bell's fourth (historical) paradigm? Or is this the way to "*biotecnocenosi*" (a state where humans and machines coexist as a part of the same intellectual entity)?

## 5. FEW OBSERVATIONS TO READ AS A PLEA.

It is no straightforward way to write the final section of an essay that aims to open up a debate. This is even clearer if we consider the last few lines of the previous paragraph. Many questions lead historians to evaluate their points of view. We may have many doubts, but, at the same time, the few certainties enable us to train historians to cope with that new phase of historical research when it is to be exclusively automated and digitised.
It is already evident that Artificial Intelligence can read and transcribe handwritten text. Still, we must expand its capability to infer meanings from the words (and what meanings it must deduce).
It is also evident the feasibility of applying computational analysis tools to mine pieces of information from detailed data.
Nevertheless, thanks to our study conducted on Biscari heritage, we acknowledge a thin line between analogue and digital research. A connected-to-the-internet scholar is not a digital historian. Those who use ITC to satisfy the analogue methodology are not out of the traditional approach. He remains an analogue historian.
A Digital Historian is *HomoLogatus*, *i.e.*, a scholar who infers analogically but analyses with a digital mind, thanks to the Machine and its computable language. Next-generation historians will live in an interconnected-symbiotically Reality where humans and technological tools have the same role in discovering the Past.